\documentclass[aps,twocolumn,showpacs,amsmath,floatfix]{revtex4}
\usepackage{graphicx}%
\usepackage{dcolumn}
\usepackage{color}
\usepackage{amsmath}
\usepackage{here}

\begin{document}

\newcommand{\be}{\begin{equation}}
\newcommand{\ee}{\end{equation}}
\newcommand{\rojo}[1]{\textcolor{red}{#1}}

\title{ Discrete solitons and nonlinear surface modes
in semi-infinite waveguide arrays}

\author{Mario I. Molina}

\affiliation{Departments de F\'{\i}sice, Faculae de Cinemas,
Universidad de Chile, Santiago, Chile}

\author{Rodrigo A. Vicencio}

\affiliation{Max-Planck-Institut fur Physik komplexer Systeme,
D-01187 Dresden, Germany}

\author{Yuri S. Kivshar}

\affiliation{Nonlinear Physics Center, Research School of Physical
Sciences and Engineering, Australian National University, Canberra
ACT 0200, Australia}

%\date{\today}

\begin{abstract}
We discuss the formation of self-trapped localized states near the
edge of a semi-infinite array of nonlinear waveguides. We study a
crossover from nonlinear surface states to discrete solitons by
analyzing the families of odd and even modes centered at different
distances from the surface, and reveal the physical mechanism of the
nonlinearity-induced stabilization of surface modes.
\end{abstract}

%\ocis{030.1640, 190.4420}

\maketitle

\newpage

Surface modes are a special type of waves localized at an
interface between two different media. Surface states have been
studied in different fields of physics, including
optics~\cite{Yeh_APL_78,tomlinson}, where such waves are confined
to the interface between periodic and homogeneous dielectric
media, and nonlinear dynamics of discrete chains~\cite{physica_d}.
In periodic systems, staggered modes localized at surfaces are
known as Tamm states~\cite{Tamm_ZPhys_32}, first found as
localized electronic states at the edge of a truncated periodic
potential.

Recently it was predicted theoretically and demonstrated
experimentally that nonlinear self-trapping of light near the edge
of a waveguide array with {\em self-focusing} nonlinearity can
lead to the formation of discrete surface
solitons~\cite{OL_surface, NLGW_nonlin_Tamm}. It was found that
the self-trapped surface modes acquire some novel properties
different from those of the discrete solitons in infinite
lattices: they can only exist above certain power level and for
the same amount of power, it is possible to have, in some
conditions, up to two surface modes, one stable and the other
unstable.

In this Letter, we reveal and explain the physical mechanism of
the nonlinearity-induced stabilization of surface modes and their
existence above a certain power threshold. In particular, we
analyze the families of odd and even modes placed at different
distances from the surface, and discuss a crossover between the
nonlinear surface states and discrete solitons of a semi-infinite
lattice.

We study a semi-infinite array of identical, weakly coupled
nonlinear optical waveguides [as shown in the inset of Fig. 1(a)]
described by the system of coupled-mode equations~\cite{CJ,book} for
the normalized mode amplitudes $E_n$,
\begin{equation}
\label{eq:1}
   \begin{array}{l} {\displaystyle
i {d E_{1}\over{d z}} + \alpha \ E_{1} +  E_{2} + \gamma\
|E_{1}|^{2} E_{1}= 0,
   } \\*[9pt] {\displaystyle
i {d E_{n}\over{d z}} + \alpha \ E_{n} +  (E_{n+1} + E_{n-1}) +
\gamma\ |E_{n}|^{2} E_{n} =0,}
\end{array}
\end{equation}
where $n \geq 2$, the propagation coordinate $z$ is normalized to
the intersite coupling $V$, $E_{n}$ are defined in terms of the
actual electric fields ${\cal E}_{n}$ as $E_{n} = (2 V \lambda_{0}
\eta_{0}/\pi n_{0} n_{2})^{1/2} {\cal E}_{n}$, where $\lambda_{0}$
is the free-space wavelength, $\eta_{0}$ is the free-space
impedance, $\alpha$ is the normalized linear propagation constant of
each waveguide, $n_{2}$ and $n_{0}$ are nonlinear and linear
refractive indices of each waveguide, and $\gamma = \pm 1$ defines
focusing or defocusing nonlinearity, respectively.

We look for stationary modes of the waveguide array in the form
$E_{n}(z) = \exp(i\beta z) E_{n}$, where $\beta$ is the
nonlinearity-induced shift of the propagation constant. For
$\gamma =0$ we use the ansatz $E_{n}\sim \sin(n k)$ and obtain the
linear spectrum $\beta= \alpha + 2 \cos \,k$, $(0\leq k \leq
\pi)$, and no localized surface modes. The presence of
nonlinearity in the model (1) can give rise to {\em new localized
states}. To find those modes, we analyze the stationary equations
(1) where, without loss of generality, we scale out the parameter
$\alpha$.

%%%%%%%%%%%%%%%%%%%%%fig1
\begin{figure}
\centerline{\includegraphics[width=3.2in]{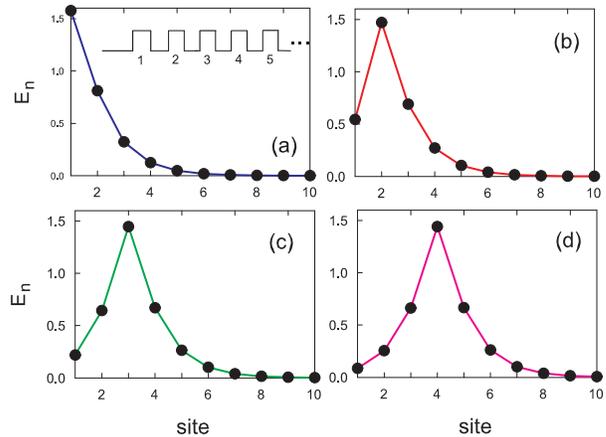}}
\caption{Examples of surface localized modes at $\beta =3$ in the
array of focusing waveguides ($\gamma =+1$) centered at different
distances $d=0,1,2,3$ from the array edge. }
\label{fig1}
\end{figure}
%%%%%%%%%%%%%%%%%%%%

For given $\beta$, the system of stationary equations is solved
numerically by a multi-dimensional Newton-Raphson scheme. Since we
are interested in surface localized modes, we look for the states
with maxima near the surface that decay quickly away from the array
edge. Similar to an infinite array, these states could be centered
at a waveguide site, or centered between waveguides. In an infinite
discrete chain, such modes are known as {\em odd} and {\em even}
states, respectively. In our calculations, we take $N = 51$
waveguides and explore both focusing and defocusing nonlinearities
looking for localized modes below and above the linear spectrum
band, $|\beta|< 2$.

Figures 1(a-d) and 2(a-d) show examples of the nonlinear localized
states centered at different sites near the surface, for both
focusing ($\gamma=+1$, $\beta=3$) and defocusing ($\gamma=-1$,
$\beta=-3$) nonlinearities, respectively. The surface state centered
at the site $n=1$ and shown in Fig. 1(a) was predicted earlier by
Markis {\em et al.}~\cite{OL_surface}. The existence of {\em
multiple localized states} near the surface and their stability are
important characteristics of an interplay between nonlinearity and
discreteness of the array, on one hand, and the surface created by
the lattice truncation, on the other. In both the cases, the states
(b,c) describe a crossover regime between the modes (a) with the
maximum amplitude at the surface and the modes (d) which are weakly
affected by the presence of the surface.

\begin{figure}[t]
\includegraphics[width=3.2in]{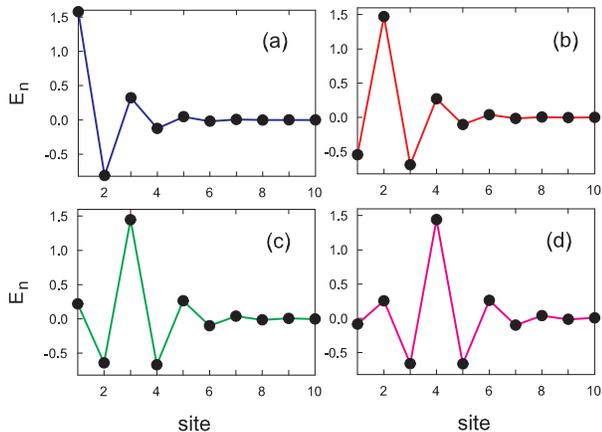}
\caption{Examples of localized surface modes at $\beta =-3$ in the
array of defocusing waveguides ($\gamma =-1$) located at different
distances $d=0,1,2,3$ from the array edge.} \label{fig2}
\end{figure}

To analyze the linear stability of each nonlinear stationary state
found numerically, we introduce a weak perturbation as $E_{n}(z) =
E_{n} + [u_{n}(z) + i v_{n}(z)] \exp(i\beta z)$, and obtain linear
evolution equations for $u_{n}$ and $v_{n}$, that can be expressed
in a compact form by defining the real vectors $\delta {\bf U}
\{u_{n} \}$ and $\delta {\bf V} = \{ v_{n}\}$, and real matrices
${\bf A}=\{A_{nm}\} = \{\delta_{n,m+1} + \delta_{n,m-1} + (-\beta
+ 3 \gamma |E_{n}|^{2} )\ \delta_{n,m} \}$ and ${\bf B}=\{B_{nm}\}
= \{ \delta_{n,m+1} + \delta_{n,m-1} + (-\beta + \gamma
|E_{n}|^{2} )\ \delta_{n,m}\}$. With these definitions, the
combined linear equations can be written in the form,
$\ddot{\delta {\bf U}} + {\bf {B A}}\ \delta {\bf U} =0$,
$\ddot{\delta {\bf V}} + {\bf {A B}}\ \delta {\bf V} = 0$, where
the dot stands for the derivative in $z$. Therefore, linear
stability of nonlinear localized modes is defined by the
eigenvalue spectra of the matrices ${\bf{A B}}$ and ${\bf{B A}}$.
If any of the real eigenvalues is negative, the corresponding
nonlinear stationary solution is unstable; otherwise, the solution
is stable. Results of this analysis are consistent with the
so-called Vakhitov-Kolokolov stability criterion of nonlinear
localized modes, and the solitons determined by the slope of the
power dependence $P = \sum_{n} |E_{n}|^2$, i.e. the states with $d
P/d \beta< 0$ for $\beta> 0$ or $d P/d \beta > 0$ for $\beta<0$,
should be unstable.

\begin{figure}[t]
\includegraphics[width=3.2in]{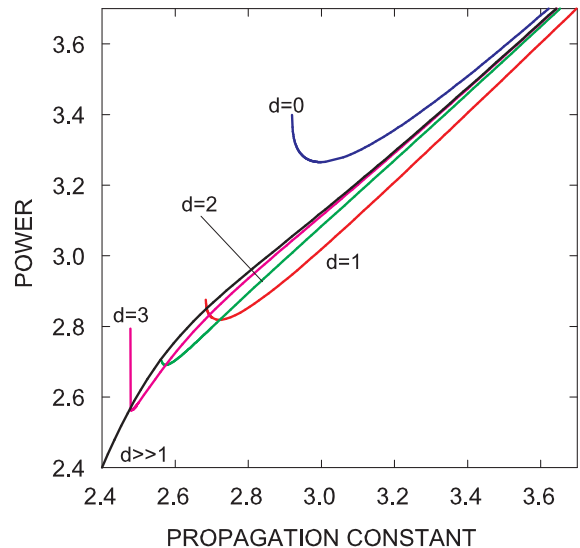}
\caption{Normalized power vs. propagation constant $\beta$ for the
surface modes shown in Fig. 1 located at different distances
$d=0,1,2,3$ from the surface. Black curve corresponds to the
discrete soliton in an infinite array.} \label{fig3}
\end{figure}

Figure 3 shows the power $P$ of the localized surface states vs. the
propagation constant for the modes in the focusing waveguides shown
in Figs. 1(a-d), and the corresponding curves for the modes of the
defocusing waveguides are mirror images. Direct numerical
simulations and stability analysis confirm the validity of the
Vakhitov-Kolokolov stability criterion; the instability region
decreases as the center of the localized mode gets shifted away from
the array edge.

Similarly, we have also found even localized modes, akin to the
modes found earlier for a semi-infinite nonlinear
lattice~\cite{physica_d}, and verified that all in-phase even modes,
for the focusing nonlinearity, and out-of-phase odd modes, for
defocusing nonlinearity are all unstable, similar to the case of an
infinite array.

In order to get a deeper insight into the physics of the nonlinear
stabilization of the surface modes, we calculate the effective
energy of the mode $H = -\sum_n (E_{n}E_{n+1}^{*} +
E_{n}^{*}E_{n+1}) -\frac{1}{2} \sum_n |E_{n}|^{4}$ as a function
of the distance of the collective coordinate of the mode $X =
P^{-1} \sum_n n|E_{n}|^2$ from the surface, similar to the case of
a defect~\cite{kivshar}. We apply a constraint method and start
from the solution centered at the site ${\bar n}$ for given values
of $\beta$ and $P$. Our goal is to obtain all intermediate
solutions between the odd and even stationary configurations for
the same power. We proceed as follows:(i) We calculate an odd
stationary mode centered at ${\bar n}$ and obtain all $\{E_{n}\}$
and the power $P$, (ii) fix the amplitude at the site ${\bar n}+1$
to be $E_{{\bar n}+1} + \epsilon$, (iii) solve the Newton-Raphson
equations for all remaining $E_{m}$ ($m\neq {\bar n}+1$) with the
constraint that the power be kept at $P$, arriving at an
intermediate state centered between ${\bar n}$ and ${\bar n}+1$,
and finally (iv) vary $\epsilon$ and repeat the procedure until we
reach the even configuration, where the amplitudes at the sites
${\bar n}$ and ${\bar n}+1$ coincide.

\begin{figure}[t]
\includegraphics[width=1.6in]{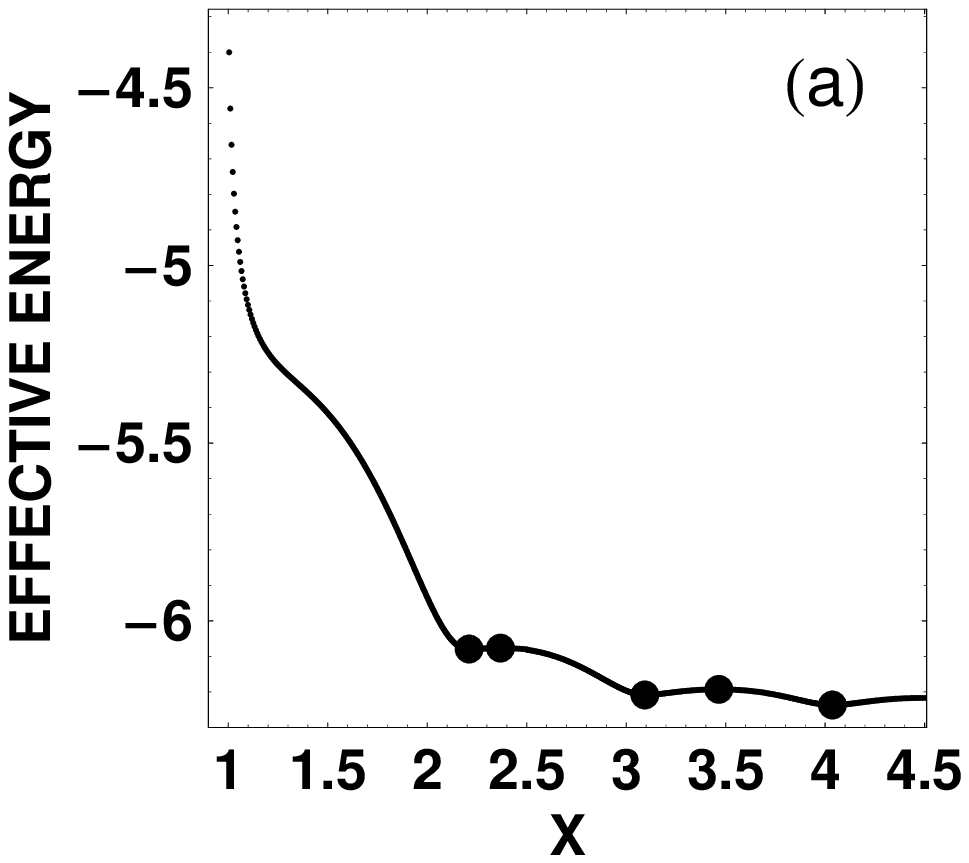}
\includegraphics[width=1.6in]{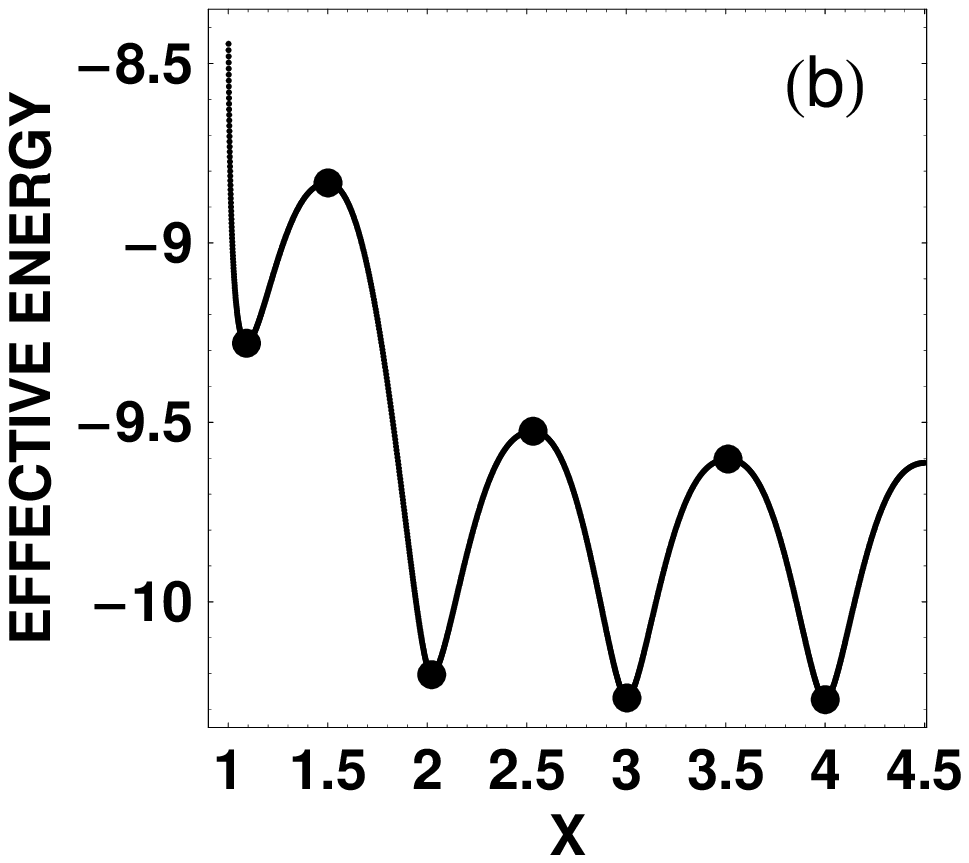}
\caption{Effective energy of surface modes vs. coordinate $X$ near
the edge of the array: (a) below ($P=2.85$) and (b) above
$(P=4.05$) threshold. Black dots correspond to the stationary
solutions found without constraint.
 } \label{fig4}
\end{figure}

In Figs.~4(a,b), we show the effective energy of a surface
localized mode in a semi-infinite array, $U_{\rm eff}(X) \equiv
H(X)$, calculated for two different power values. The extremal
points of this curve defined by the condition $dH/dX=0$ correspond
to the stationary localized solutions in the system.

In comparison with an infinite array, the truncation of the
waveguide array introduces an effective {\em repulsive} potential,
that is combined with the periodic (Peierls-Nabarro) potential of
an infinite waveguide array. As a result, discrete surface modes
are possible neither in the linear regime nor in the continuous
limit. As we see from Fig.~4(a), for low powers there exists no
solution of the equation $dH/dX=0$ at the surface site $n=1$; this
corresponds to the fact that no surface state is found below the
power threshold~\cite{OL_surface}. However, the modes localized at
the sites $n \geq 2$ are still possible.

If the power exceeds the threshold $P=3.26$, discreteness overcomes
a repulsive force of the surface and the surface localized state
becomes possible, as shown in Fig.~4(b). The correspondence between
the stationary solutions found without constraints (black dots) and
the solutions obtained as extremal points using the constraint
methods is perfect. As expected, all odd modes are stable compared
to even modes, and they all correspond to the condition $dH/dX = 0$.

\begin{figure}[t]
\includegraphics[width=3.2in]{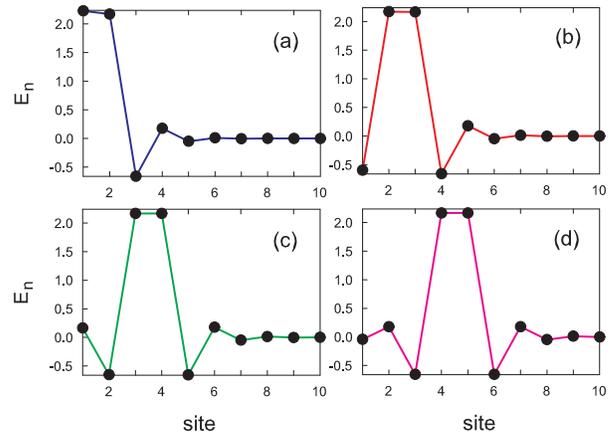}
\caption{Examples of stable flat-top localized surface modes at
$\beta =-4$ in the array of defocusing waveguides ($\gamma =-1$)
centered between different sites near the edge.} \label{fig5}
\end{figure}

We also found many other discrete surface modes, including the
so-called flat-top surface modes that generalize the corresponding
modes of infinite chains~\cite{darm}, and  two-soliton bound
states or surface twisted modes, which are stable below a certain
threshold in the propagation constant. Examples of flat-top modes
for defocusing nonlinearity are shown in Fig.~5 for $\beta =-4$,
and their stability is defined by the Vakhitov-Kolokolov
criterion.

In conclusion, we have analyzed different types of nonlinear
localized modes near the edge of a semi-infinite waveguide array
and revealed the mechanism of the nonlinearity-induced
stabilization and power threshold. In addition, we have
demonstrated that a similar approach can be applied to other types
of nonlinear discrete surface modes, such as flat-top modes and
twisted modes, as well as to the case of staggered modes in
defocusing waveguides.

This work has been supported by Conicyt and Fondecyt grants
1050193 and 7050173 in Chile, and in part by the Australian
Research Council in Australia.

\end{document}